\documentclass[aps,twocolumn,superscriptaddress,showpacs]{revtex4}
\usepackage{amsmath}
\usepackage{graphicx}

\begin{document}
% You should use BibTeX and revtex.bst for references
\bibliographystyle{apsrev}

%\preprint{}

\title{Detecting Spin-Polarized Currents in Ballistic Nanostructures }

\author{R. M. Potok}
\affiliation{Department of Physics, Harvard University, Cambridge, 
Massachusetts 02138}

\author{J. A. Folk}
\affiliation{Department of Physics, Harvard University, Cambridge, 
Massachusetts 02138}
\affiliation{Department of Physics, Stanford University, Stanford, 
California 94305}

\author{C. M. Marcus}
\affiliation{Department of Physics, Harvard University, Cambridge, 
Massachusetts 02138}

\author{V. Umansky}
\affiliation{Braun Center for Submicron Research, Weizmann Institute 
of Science, Rehovot 76100, Israel}

\date{6/12/02}

\begin{abstract}We demonstrate a mesoscopic spin polarizer/analyzer 
system that allows the spin polarization of current
from a quantum point contact in an in-plane magnetic field to be 
measured.  A transverse focusing geometry
is used to couple current from an emitter point contact into a 
collector point contact.  At large in-plane fields, with the
point contacts biased to transmit only a single spin ($g < e^2/h$), 
the voltage across the collector depends on the spin
polarization of the current incident on it.  Spin polarizations of
$> 80 \%$ are found for both emitter and collector at $300~mK$ and 
$7~T$ in-plane field.
\end{abstract}

\maketitle

The detection of single electron spins has been the aim of extensive 
experimental efforts for many years.  In addition to
providing a new tool to investigate the physics of mesoscopic 
devices, there is hope that the ability to manipulate and
measure electron spins in a solid state system may open the way for 
quantum information processing \cite{Loss,
Golovach}.  However, the long coherence times
\cite{Kikkawa} that make electron spins such a promising system for 
quantum manipulation result fundamentally from
their weak coupling to the environment,  and this makes the task of 
measuring spin difficult.

In this Letter we demonstrate a technique to measure spin 
by converting the problem into the easier one of
measuring charge.    At low field and low temperature, a narrow 
constriction in a 2D electron gas (2DEG), known as a
quantum point contact (QPC) [see Fig.~1(a)], transmits through two 
spin-degenerate channels, producing conductance
plateaus at integer multiples of
$2e^2/h$.    When a large in-plane magnetic field is applied, the 
degeneracy is lifted and conductance becomes quantized
in multiples of
$1e^2/h$ [Fig.~1(b)] \cite{Wharam, vanWees}.  While it is widely 
believed that the
$e^2/h$ plateau is associated with spin-polarized transmission, this 
has not been established experimentally to our
knowledge.  One key result of this Letter is the demonstration that 
point contacts do operate as spin emitters and detectors,
and therefore allow the detection of 
spin polarization to be accomplished by simply measuring electrical resistance.

\begin{figure}
    \label{fig1}
    \includegraphics[width=3.25in]{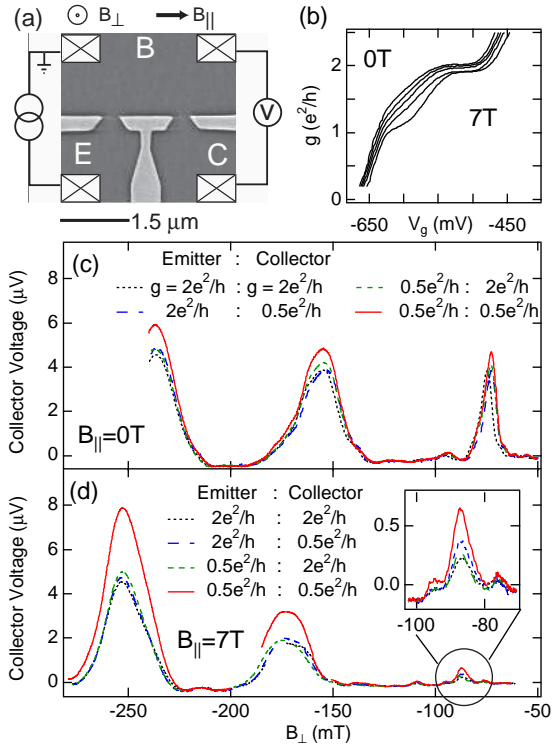}
    \caption{\small{(a) SEM micrograph of a device similar to the one measured in 
this experiment, two quantum
point contacts in a transverse focusing geometry with perpendicular 
($B_{\bot}$) and in-plane
($B_\parallel$) magnetic fields oriented as shown.  With a fixed 
current applied between emitter (E) and
base (B), the voltage between base and collector (C) showed focusing 
peaks as a function of
$B_{\bot}$. (b) At
$T=300~mK$, both  point contacts showed conductance
quantized in units of
$2e^2/h$ at
$B_\parallel=0$, and in units of $e^2/h$ at large $B_\parallel$.  (c) At $B_\parallel=0$,
the collector voltage was nearly  independent of the conductances
of the two point contacts. (d) At $B_\parallel=7~T$ the focusing 
peaks were enhanced only when both
emitter and collector were set to
$g=0.5e^2/h$. The enhancement demonstrates that both emitter and 
detector are spin selective, by
Eq.~(1).}}
    \end{figure}

Our experiment is based on a technique known as transverse electron focusing
\cite{vanHouten}, which has been used previously to study phenomena 
ranging from anisotropy in the band structure of
metals
\cite{Sharvin, Tsoi} and semiconductors \cite{Goldoni, Ohtsuka} to 
composite fermions in the fractional quantum Hall
regime \cite{Goldman}.  This device geometry [Fig. 1(a)] allows 
electrons from a spin-polarizing emitter---in this case a
QPC---to be coupled into a second QPC serving as a spin-sensitive collector. A magnetic
field, $B_\perp$, applied perpendicular to the 2DEG plane, bends and focuses ballistic
electron trajectories from the emitter to the collector, resulting in peaks in the
base-collector voltage [Figs.~1(c) and
1(d)] whenever the spacing between point contacts is an integer multiple of the cyclotron
diameter,
$2m^*v_F/eB_\perp$, where $m^*$ is the effective electron mass and $v_F$ the Fermi
velocity.

The coupling efficiency between emitter and collector can be quite 
high in clean 2DEG materials, allowing the two QPCs to
be separated by several microns.  This separation is 
useful for investigating spin physics in mesoscopic structures
because it allows spin measurements to be decoupled from the device 
under test, simplifying the interpretation of
results.  A further advantage of a focusing geometry is that only 
ballistic trajectories contribute to the signal, so spin
detection occurs very quickly ($ < 10~ps$) after the polarized 
electrons are emitted, leaving little time for spin relaxation.

In the present experiment, the focusing signal is measured as a voltage 
between collector and base regions, with fixed current
applied between emitter and base [Fig.~1(a)].  With the collector 
configured as a voltage probe, current injected
ballistically into the collector region at the focusing condition 
must flow back into the base region, giving rise to a voltage
$V_{c}=I_{c}/g_{c}$ between  collector  and base, where $I_{c}$ is 
the current injected into the collector and
$g_{c}$ is the conductance of the collector point contact.  For this experiment both point
contacts are kept at or below one channel of conductance; therefore the collector voltage
may be written in terms of the  transmission of the collector point contact, $T_c$
($\le 1$), as
$V_c=(2e^2/h)^{-1} I_c/T_c$.

To analyze how spin polarization affects the base-collector voltage, we assume
$I_{c}\propto I_{e} T_{c}$, where
$I_e$ is the emitter current, and the constant of proportionality 
does not depend on the transmissions of either of the
point contacts.  In the absence of spin effects, one then expects
$V_{c}$ to be independent of
$g_{c}$. Because
$I_e$ is fixed, $V_{c}$ would also be independent of the 
emitter conductance,
$g_{e}$.

Taking into account different transmissions for the two spin channels, however,
one expects the voltage on the collector to double if both emitter and collector
pass the same spin,  or drop to zero if the two pass opposite
spins.  This conclusion assumes that a spin polarized current 
injected into the collector region will lose all polarization
before flowing out again.  Under these conditions, the collector 
voltage generally depends on the polarization of the
emitter current
$P_e=(I_{\uparrow} - I_{\downarrow})/(I_{\uparrow} + I_{\downarrow})$ 
and the spin selectivity of the collector
$P_c=(T_{\uparrow}-T_{\downarrow})/(T_{\uparrow}+T_{\downarrow})$ in 
the following simple way
\cite{me}:
\begin{equation} V_{c}\propto\frac{h}{2e^2}I_{e}(1+P_eP_c).
\end{equation} Note from Eq.~1 that colinear and complete spin 
polarization ($P_e = 1$) and spin selectivity ($P_c =
1$)  gives a collector voltage twice as large as when 
{\em either} emitter or collector is not spin
polarized.

The focusing device was fabricated on a high-mobility two-dimensional 
electron gas (2DEG) formed at the interface of a
$\rm GaAs/Al_{0.36}Ga_{0.64}As$ heterostructure, defined using Cr/Au surface 
depletion gates patterned by
  electron-beam lithography, and contacted with nonmagnetic (PtAuGe) ohmic 
contacts. The 2DEG was
$68~nm$ from the Si delta-doped layer ($n_{Si}= 2.5 \times 10^{12} 
~cm^{-2}$) and
$102~nm$ below the wafer surface.  Mobility of the unpatterned 2DEG was
$5.5\times10^6~cm^2/Vs$ in the dark, limited mostly by remote 
impurity scattering in the relatively shallow structure,
with an estimated background impurity level $< 
5\times10^{13}~cm^{-3}$.  With an electron density of
$\sim 1.3 \times 10^{11}~cm^{-2}$, the transport mean free path was
$\sim 45~\mu m$, much greater than the distance ($1.5~\mu m$) between 
emitter and collector point contacts.  The
Fermi velocity associated with this density is $v_F=2\times 
10^7~cm/s$, consistent with the observed $\sim 80~mT$
spacing between focusing peaks.

Measurements were performed in a $^3He$ cryostat with a base 
temperature of $300~ mK$.  A conventional
superconducting solenoid was used to generate in-plane fields,
$B_\parallel$, and a smaller superconducting coil wound on the 
refrigerator vacuum can allowed fine control of the
perpendicular field, $B_\perp$
\cite{Folk01}.  $B_\parallel$ was oriented along the axis between
  the two point contacts, as shown in Fig.~1(a).

Independent ac current biases of $1~nA$ were applied between base and 
emitter ($17~Hz$), and base and collector
($43~Hz$), allowing simultaneous lock-in measurement of the emitter 
conductance (base-emitter voltage at
$17~Hz$),
  collector conductance (base-collector voltage at $43~Hz$), and the 
focusing signal (base-collector voltage at
$17~Hz$).  The base-collector current bias was found to have no 
effect on the focusing signal.  Additionally, the focusing
signal was found to be linear in base-emitter current for the small currents used in this
measurement.

Measurements  were taken over several thermal cycles of the device.
While details of focusing peak shapes and point contact conductance 
traces changed somewhat upon thermal cycling, their
qualitative behavior did not change.  Although all of the data presented in this paper
comes from a single device, the results were confirmed in a similar device on the same
heterostructure.

Spin polarization and spin selectivity of the point contacts were 
detected by comparing the focusing signal (the
collector voltage at the top of a focusing peak) for various 
conductances of the emitter and collector point
contacts.  At $B_\parallel=0$, where no static spin polarization is 
expected, the focusing signal was found to be nearly
independent of the conductances of both emitter and collector point 
contacts, as shown in Fig.~1(c). In contrast, at
$B_\parallel=7~T$, the focusing signal observed when both the emitter 
and collector point contacts were set well below
$2e^2/h$ was larger by a factor of $\sim 2$ compared to the signal 
when either emitter or collector was set to
$2e^2/h$, as seen in Fig.~1(d). A factor-of-two enhancement is 
consistent with Eq.~(1) for fully spin polarized emission
and aligned, fully spin-selective detection.

To normalize for overall variations in transmission through the bulk from the emitter to 
the collector (for instance upon thermal cycling),  the
focusing signal at any emitter or collector setting can be normalized
by the value  when both the emitter and collector are set to
$2e^2/h$. We denote the point contact settings as $(x:y)$ where $x$ 
is the conductance of the emitter and $y$ is the
conductance of the collector, both in units of $e^2/h$. For instance,
$(2:2)$ indicates  both emitter and collector set to
$2e^2/h$  (expected to be unpolarized in any field), while 
$(0.5:0.5)$ indicates both point contacts set to
$0.5e^2/h$ (expected to be polarized in a sizable in-plane field). 
Ratios are then denoted $(x: y)/(2:2)$.

\begin{figure}
    \label{fig2}
    \includegraphics[width=3.25in]{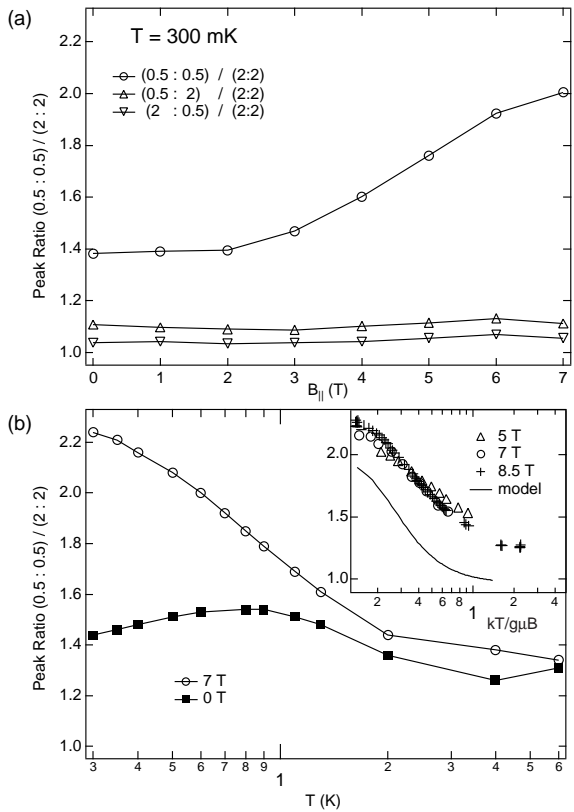}
    \caption{\small{(a) The height of the third 
focusing peak as a function of $B_\parallel$ for different conductances of the point contacts
$(x:y)$, where
$x$ is the emitter conductance and $y$ is the collector conductance (in units of 
$e^2/h$), all normalized by the
$(2:2)$ focusing peak height.  According to Eq.~(1), a factor of two in 
the ratio indicates fully spin
polarized emission and detection.  (b)  Temperature dependence of 
the ratio of focusing signals
$(0.5:0.5)/(2:2)$ for $B_\parallel=7~T$ and $0~T$. (a) and (b) are from different
cooldowns.  Inset: Ratio
$(0.5:0.5)/(2:2)$ for
$B_\parallel = 5, 7,$ and 
$8.5~T$ plotted as a function of 
$kT/g\mu B_\parallel$.  The solid curve is the prediction of a simple 
model (see text) that  accounts for only
thermal broadening in the leads.}}
    \end{figure}

Figures 2 and 3 show the focusing signal ratios for the third focusing 
peak ($B_\perp \sim
230-250~mT$), chosen because its height and structure in the
$(2:2)$ condition were less sensitive to  $B_\parallel$ and small 
variations in point contact tuning compared to the first
and second peaks. However,  all peaks showed qualitatively similar behavior.

Figure~2(a) shows that only the ratio
$(0.5:0.5)/(2:2)$ grows with $B_\parallel$, reaching a value $\sim 2$ 
at $7T$, while the other ratios,
$(2:0.5)/(2:2)$ and
$(0.5:2)/(2:2)$, are essentially independent of in-plane field, as 
expected from Eq.~(1) if no spin selectivity exists when the conductance is $2~e^2/h$. At
$B_\parallel = 0$, we find
$(0.5:0.5)/(2:2) \sim 1.4$, rather than the expected
$1.0$, for this particular cooldown.  As discussed below, these ratios 
fluctuate somewhat between thermal cycles.

Temperature dependences of the $(0.5:0.5)/(2:2)$ ratio are shown in Fig.~2(b) for a
different cooldown.  At 
$B_\parallel = 7T$, the ratio
$(0.5:0.5)/(2:2)$ decreases from $\sim2.2$ at
$T=300~mK$ to the zero-field value of $1.4$ above $2K$. Note that 
$2K$ is roughly the temperature at which
$g
\mu B_\parallel/kT \sim 1$, using the GaAs g-factor $g =-0.44$.  At 
$B_\parallel = 0$, the ratio $(0.5:0.5)/(2:2)$
remains near
$1.4$, with only a weak temperature dependence up to
$6K$.  

The inset of Fig.~2(b) shows that focusing data at different values of $B_{\parallel}$ 
scale to a single curve when plotted as a function of
$kT/g\mu B_\parallel$, suggesting that both spin-polarized emission 
and spin-selective detection arise from an energy
splitting that is linear in
$B_\parallel$. A simple model that accounts roughly for the observed scaling 
of the focusing signal assumes that the point contact 
transmission, $T(E)$, is  $0$ for $E< E_0$, and
$1$ for $E> E_0$, where $E$ is the electron kinetic energy and  $E_0$ is a
gate-voltage-dependent threshold.  Spin selectivity then results from 
the Zeeman splitting of the two spin sub-bands, and
is reduced by thermal broadening.  Except for a vertical offset of
$\sim 0.4$, this simple model agrees reasonably well with the data 
[Fig.~2(b), inset].

Fig.~3(a) shows the evolution of spin selectivity in the collector point contact as
a function of its conductance.  At
$B_\parallel=6~T$, with the emitter point contact set to $0.5 e^2/h$, 
the collector point contact is swept from  $2
e^2/h$ to 0.  The focusing signal increases as the collector point 
contact conductance is reduced below  $2 e^2/h$,
saturating as the collector conductance goes below the
$e^2/h$ spin-split plateau.  The polarization saturates completely 
only well into the tunneling regime, below
$\sim0.5 e^2/h$.  Similar to the effect seen in Fig.~2(b), spin 
selectivity decreases with increasing temperature,
approaching the zero field curve at
$1.3~K$.

\begin{figure}
    \label{fig3}
    \includegraphics[width=3.25in]{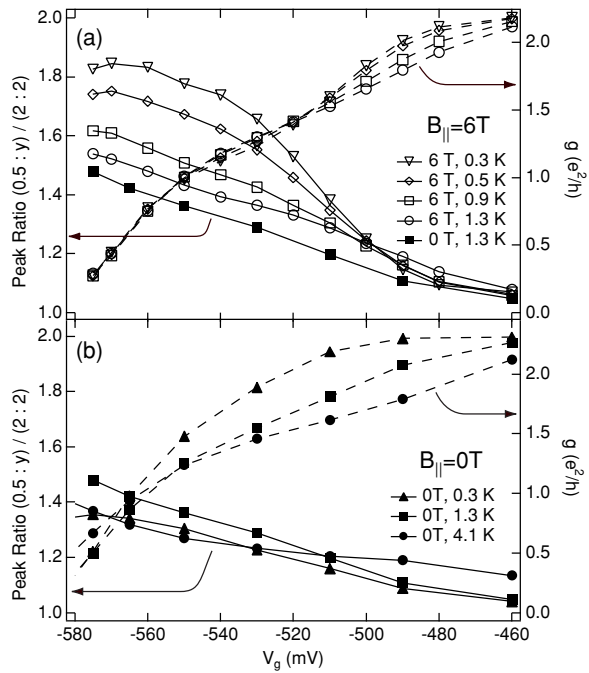}
    \caption{\small{(a) Focusing signal ratio $(0.5:y)/(2:2)$ and collector conductance
$g$ at 
$B_\parallel=6~T$ as a function of the
voltage applied to one of the collector gates, with the emitter fixed at
$g=0.5e^2/h$. This shows the onset of spin selectivity as the 
collector point contact is brought into the
tunneling regime, $g< 2e^2/h$. (b) The same data taken at
$B_\parallel=0$, showing little temperature dependence up to
$4~K$.  A mild 0.7 structure in the conductance becomes more prominent at
$1.3~K$.}}
    \end{figure}

Fig.~3(b) shows the same measurement taken at $B_\parallel=0$.  The 
focusing peak rises slightly when both point
contacts are set below one spin degenerate channel.  Unlike at high 
field, however, the increase of the focusing signal is
very gradual as the point contact is pinched off.  In addition, 
temperature has only a weak effect.

As mentioned above, both the low and high field ratios $(0.5 : 0.5)/(2:2)$ were measured
to be larger than their ideal theoretical values of 1 and 2 respectively. Sampled over
multiple thermal cycles, several gate voltage settings (shifting the point contact 
centers by $\sim 100~nm$), and different focusing peaks, the ratio
at $B_{\parallel}=0$ varied between 1.0 and 
1.6, with an average value of 1.25 and a standard deviation $\sigma= 0.2$. The average
value of the ratio at $B_\parallel=7~T$ was 2.1, with $\sigma=0.1$.

Both point contacts display a modest amount of zero-field 0.7 
structure \cite{Thomas, Cronenwett}, as seen in Figs.~1(b)
and 3(b).  Although a static spin polarization associated with 0.7 
structure would be consistent with our larger-than-one
ratio $(0.5:0.5)/(2:2)$ at zero field, this does not explain the
enhanced ratio found {\em both} at zero field and high field.
Rather, we believe the enhancement is due to a slight increase in the 
efficiency of focusing for $(T_c, T_e) < 1$.  For example,
more of the emitted current may be focused into the collector as the 
point contacts are pinched off, causing deviations from
the assumption $I_c\propto I_e T_c$.  This explanation is also 
consistent with the weak temperature dependence of the
zero-field ratio up to $4~K$, which would not occur if the 
enhancement were due a static polarization at zero
field.

An unexplained feature of our data is the relative suppression of the 
lower-index focusing peaks---particularly the first peak---in a large in-plane field, as
seen in Figs.~1(c) and 1(d).  This effect was observed over multiple thermal cycles and
for all point  contact positions.  The effect is not readily
explained as a field-dependent change in the scattering rate, as 
neither the bulk mobility, nor the width of the focusing peak is
affected.  Also, the effect is not obviously related to spin, as it 
occurred for both polarized and unpolarized
point contacts.

In conclusion, we have developed a new method for creating and remotely 
detecting spin currents using quantum point contacts.
The technique has allowed a first demonstration of what was widely expected, namely that a
point contact in an in-plane field can act as a spin polarized emitter and a spin
sensitive detector.  From our perspective, however, this result also has a larger
significance: it is the first demonstration of a wholly new technique to measure
spin-current  from a mesoscopic device using a remote electrical spin detector. In future
work, this technique can be applied to more subtle  mesoscopic spin systems such as
measuring spin currents from open or Coulomb-blockaded quantum dots, or directly 
measuring spin precession due to a spin-orbit interaction.

We acknowledge valuable discussions with H. Bruus, S. Cronenwett, A. 
Johnson, and H. Lynch.  This work was supported in part by
ARO-MURI (DAAD 19-99-1-0215) and DARPA-SpinS (DAAD 19-01-1-0659).  JAF acknowledges partial
support from the Stanford Graduate  Fellowship; RMP acknowledges support as an ARO Graduate
Research Fellow.

\vspace{0.1in}
% Create the reference section using BibTeX
%\bibliography{josh}
\small{
 }

\end{document}